 \documentclass[preprint2]{aastex}

\newcommand{\Ell}{E_\parallel}      
\newcommand{\rhoGJ}{\rho_{{\rm GJ}}}  
\newcommand{\rlc}{\varpi_{\rm LC}} 
\newcommand{\inc}{\alpha_{\rm i}}  

\shorttitle{High Energy Emission from Pulsar Magnetospheres}
\shortauthors{Hirotani}

\begin{document}


\title{Outer-gap vs. Slot-gap Models for Pulsar High Energy Emissions:
       The Case of the Crab Pulsar}


\author{Kouichi Hirotani}
\affil{ASIAA/National Tsing Hua University - TIARA,\\
       PO Box 23-141, Taipei, Taiwan\footnote{
            Postal address: 
            TIARA, Department of Physics, 
            National Tsing Hua University,
            101, Sec. 2, Kuang Fu Rd.,Hsinchu, Taiwan 300}
      }
\email{hirotani@tiara.sinica.edu.tw}




\begin{abstract}
We analytically examine the capabilities of rotation-powered pulsars
as the sources of gamma-rays 
and show that their phase-averaged gamma-ray flux is proportional to
the product of the spin-down flux and the gap 
trans-field thickness cubed irrespective of the emission models.
Applying the scheme to the Crab pulsar,
we demonstrate that the outer-gap model reproduces
the observed GeV fluxes and that
the slot-gap model reproduces at most twenty per cent of 
the observed fluxes because of the small trans-field thickness.
An implication on the relationship
between the gamma-ray and the spin-down fluxes is discussed.
\end{abstract}





\section{Introduction}
\label{sec:intro}
The launch of 
Fermi Gamma-ray Space Telescope
will soon open a new era for the studies of 
rotation-powered pulsars.
The unprecedented sensitivity and spectral resolution of the
Large Area Telescope (LAT) 
aboard Fermi Space Telescope 
will allow for detailed studies of particle acceleration and 
radiation in rotating neutron-star (NS) magnetospheres.
To make the best use of the power of LAT observations,
we need the most sophisticated model with minimum assumptions.

In all the pulsar emission models, 
$e^-$'s and/or $e^+$'s are accelerated 
by the magnetic-field-aligned electric field, $\Ell$,
to radiate photons in the open zone 
(fig.~1 in Hirotani 2008, hereafter H08)
mainly via synchro-curvature process.
In polar-cap (PC) models, emission takes place
within several NS radii above a PC surface
(Arons \& Scharlemann 1979; Daugherty \& Harding 1982, 1996).
However, such a low-altitude emission predicts 
too small beam size to produce the observed wide pulse profiles.
Therefore, extending the original idea by Arons (1983),
Muslimov and Harding (2003, hereafter MH03; 2004a, b) and
Dyks et al.~(2004, hereafter DHR04) 
sought the possibility of a wide hollow cone 
of high-energy radiation due to the flaring of field lines. 
They proposed the slot-gap (SG) model,
in which emission takes place along the last-open field lines.
Recently, Harding et al. (2008, hereafter HSDF08) demonstrated that 
the SG model reasonably reproduces
the Crab pulsar phase-resolved spectrum 
(Fierro et al.~1998; Kuiper et al.~2001; Nolan et al.~1993).

The outer-gap (OG) model gives an alternative possibility
(Cheng et al. 1986a,b; Romani 1996; Cheng \& Zhang~1996;
 Hirotani~2006a, b, hereafter H06a, b).
It differs from the SG model in the following ways:
\,$\bullet$
An OG extends between the null surface on which
the magnetic field becomes perpendicular to the rotation axis
and the light cylinder on which the plasma co-rotational
velocity would exceed the speed of light, $c$.
A SG extends between the PC surface and the light cylinder.
\,$\bullet$
For Crab, an OG occupies more than 10\% of the open magnetic fluxes 
with a large electro-static potential drop, 
$\Delta\Psi \sim 10^{15}$~V, whereas
a SG occupies several \% of the open fluxes with 
$\Delta\Psi \sim 10^{13}$~V.
\,$\bullet$
Pair production copiously takes place in OGs, whereas
it is negligible in SGs.

The purpose here is to explore further into the two models.
We examine their common emission properties
in \S~\ref{sec:analytic},
and separately consider the OG and SG models
in \S\S~\ref{sec:OG} and \ref{sec:SG}.
\S~\ref{sec:discussion} is for discussion.

\section{Gamma-ray Flux}
\label{sec:analytic}
In this letter, we concentrate on the 
phase-{\it averaged} spectrum of pulsar magnetospheric emissions,
sacrificing the examinations of light curves and 
phase-resolved spectra.
In this context, we can neglect
the aberration of photon propagation directions and the 
time-of-flight delay due to different emission altitudes from the NS,
because these two relativistic effects 
do not change the total number of photons to be detected.
Although this thought experiment does not describe any realistic
pulsar emissions, 
it significantly reduces the calculation of photon propagations
and gives the correct phase-{\it averaged} spectrum.

To investigate the upper limit of photon fluxes,
we neglect photon absorptions. 
Then the radiative transfer equation gives 
the specific intensity 
$ I_\nu \!\approx\! 2b\varrho_{\rm c} j_\nu $,
where $b \!\approx\! \gamma^{-1}$ denotes the emission beaming angle,
$\gamma$ the Lorentz factor of $e^\pm$'s,
$\varrho_{\rm c}$ the local curvature radius
of the magnetic field line, $j_\nu$ the emission coefficient,
and $2b\varrho_{\rm c}$ 
the distance interval from which the photons are detected 
by the observer (fig.~6.2 in Rybicki \& Lightman 1979).
%
Giving the emission coefficient as
$j_\nu \approx N (dP/d\nu)/(\pi b^2)$, we obtain 
\begin{equation}
  I_\nu \approx \frac{2}{\pi} \frac{\varrho_{\rm c}}{b} 
                N \frac{dP}{d\nu},
  \label{eq:Inu2}
\end{equation}
where $N$ denotes the spatial density of $e^-$'s or $e^+$'s, and
$dP/d\nu$ the radiation power per particle. 


At each magnetic azimuth $\varphi_\ast$ on the PC surface, 
we parameterize the field-line footpoint with their magnetic
colatitude $\theta_\ast$ measured from the magnetic axis.
We define that the primary photons are emitted only 
along the field lines threading the PC surface with
$\theta_\ast^{\rm min}<\theta_\ast<\theta_\ast^{\rm max}$,
where the upper boundary and the last-open field lines 
are designated with $\theta_\ast^{\rm min}$ and
$\theta_\ast^{\rm max}$, respectively.
We further introduce the dimensionless gap trans-field thickness,
$h_{\rm m} \equiv (\theta_\ast^{\rm max}-\theta_\ast^{\rm min})
 /\theta_\ast^{\rm max}$.
In this letter, 
we assume that $h_{\rm m}$ is constant for both $\varphi_\ast$ and 
$s$ (distance along the field line) for simplicity.

Let us introduce the gap meridional thickness $\Delta z$ that 
represents the distance between the last-open field line 
and the upper boundary measured perpendicularly to the field line.
Then, the observer detects emissions from the magnetic-flux cross
section $\Delta A \approx \Delta z \times 2b r\sin\theta$,
where $r\sin\theta$ refers to the distance from the magnetic axis,
and $2b$ the {\it azimuthal} full opening angle of the points
from which the photons propagate towards the observer.
For an aligned rotator, the azimuthal length
from which the photons propagate towards the observer
is given by $2b r\sin\theta$;
thus, $\Delta A=\Delta z \times 2b r\sin\theta$ holds.
For an oblique rotator, in the outer magnetosphere,
toroidal expansion of the field line flux tubes
is similar to an aligned case;
thus, $\Delta A \!\approx\! \Delta z \times 2b r\sin\theta$ 
approximately holds.
Since
$\Delta z \approx 2h_{\rm m} \rlc (r/\rlc)^2
                  \csc\theta (1+3\cos^2\theta)^{-1/2}$
holds for a dipole field, we obtain
\begin{equation}
  \Delta A \approx 2b h_{\rm m} \rlc^2 (B_\ast/B) 
                   (r_\ast/\rlc)^3,
  \label{eq:DelA2}
\end{equation}
where $r$ denotes the distance from the NS center,
$\theta$ the magnetic colatitude,
$\rlc \equiv c/\Omega$ the light cylinder radius, 
$\Omega$ the NS angular frequency.
The magnetic field strength $B$ is evaluated at the 
emission point and related with its surface value $B_\ast$ by
$B/B\ast= (r_\ast/r)^3 \sqrt{1+3\cos^2\theta}/2$.

The photon energy flux can be computed by
\begin{equation}
  \nu F_\nu = \nu I_\nu (\Delta A/d^2), 
  \label{eq:nuFnu}
\end{equation}
where $d$ is the distance to the pulsar.
Substituting equations~(\ref{eq:Inu2}) and (\ref{eq:DelA2}) 
into equation~(\ref{eq:nuFnu}), we obtain 
\begin{equation}
  \nu F_\nu \approx 
  \frac{4}{\pi} h_{\rm m}
  \frac{\Omega B_\ast}{2\pi ce} n
  \varrho_{\rm c} \nu \frac{dP}{d\nu} 
  \left(\frac{r_\ast}{d}\right)^2 
  \frac{r_\ast}{\rlc}, 
  \label{eq:nuFnu2}
\end{equation}
where $e$ is the charge on the positron;
the dimensionless particle density per magnetic flux tube, 
$n \equiv (2\pi ce/\Omega B)N$, becomes approximately $\cos\inc$,
where $\inc$ is the magnetic inclination with
respect to the rotation axis.
The quantity $\Omega B_\ast/(2\pi ce)$ expresses the typical
Goldreich-Julian (GJ) particle number density at the PC surface
(Goldreich \& Julian~1969).
It follows from equation~(\ref{eq:nuFnu2})
that the photon flux $\nu F_\nu$ does not depend on $b$.

For saturated $e^-$'s or $e^+$'s, electrostatic force balance,
$e\Ell=2e^2\gamma^4/(3\varrho_{\rm c}^2)$, 
gives the terminal Lorentz factor
\begin{equation}
  \gamma
  = 2.20 \times 10^5
    \Omega_2^{-1/2}
    \left(\frac{\varrho_{\rm c}}{0.5\rlc}\right)^{1/2}
    \left(\frac{\Ell}{\mbox{V\,m}^{-1}}\right)^{1/4},
  \label{eq:termLf}
\end{equation}
where $\Omega_2 \equiv \Omega/(10^{\,2}\mbox{\,rad\,s}^{-1})$
and $\Ell \equiv -(\mbox{\boldmath$B$}/B)\cdot \nabla \Psi$.
The non-corotational potential $\Psi$
is given by the inhomogeneous part of the Maxwell equations,
\begin{equation}
  -\nabla^2 \Psi= 4\pi(\rho-\rhoGJ)
                = (2\Omega B_z/c)\kappa,
  \label{eq:Poisson}
\end{equation}
where $\rho$ and $\rhoGJ$ denote the real and GJ charge densities,
respectively, and $B_z$ the magnetic field component 
projected along the rotation axis.
The effective charge density, $\rhoGJ-\rho$,
is parameterized by $\kappa$, 
which is a function of position.
In the Newtonian limit, we would have
$\kappa=1$ for a vacuum gap while $\kappa<1$ for a non-vacuum gap.

Since $e^\pm$'s are ultra-relativistic in the gap, 
the primary emission is dominated by the pure-curvature component. 
Thus,
\begin{equation}
  \nu \frac{dP}{d\nu}
  = \frac{3\sqrt{3}}{4\pi} \frac{ce^2\gamma^4}{\varrho_{\rm c}^2}
    x^2\int_x^\infty K_{5/3}(\xi)d\xi
  \label{eq:dPdnu}
\end{equation}
peaks at $x \equiv \nu/\nu_{\rm c}= 1.318$ with the maximum value
$x^2\int_x^\infty K_{5/3}(\xi)d\xi= 0.6826$,
where $K_{5/3}$ is the modified Bessel function of 5/3 order, and
$\nu_{\rm c}\equiv (3/4\pi)(c\gamma^3/\varrho_{\rm c})$ 
the characteristic frequency of the emission.
The electrostatic force balance,
$(2/3)e^2\gamma^4/\varrho_{\rm c}^2=e\Ell$, gives
\begin{equation}
  \nu \frac{dP}{d\nu}
  = \frac{9\sqrt{3}}{8\pi}ce\Ell 
    x^2 \int_x^\infty K_{5/3}(\xi)d\xi.
  \label{eq:dPdnu2}
\end{equation}
For a thin gap ($h_{\rm m}\!\ll\! 1$), equation~(68) in H06a gives
\begin{equation}
  \Ell \approx \frac{h_{\rm m}^2}{4}
               \left(\frac{r_\ast}{\rlc}\right)^3
               B_\ast
               \frac{\partial(-\kappa B_z/B)}{\partial(s/\rlc)},
  \label{eq:Ell}
\end{equation}
where $s$ is the distance along the field line.
For the field lines curving away (or toward) the rotation axis,
$1.0 < -\partial(B_z/B)/\partial(s/\rlc) < 1.5$
(or $1.5 < -\partial(B_z/B)/\partial(s/\rlc) < 2.2$) 
is the typical range.

Substituting $B_\ast= 2\mu/r_\ast^3$ and 
$x^2\int_x^\infty K_{5/3}(\xi)d\xi= 0.68$,
and combining equations~(\ref{eq:nuFnu2}), (\ref{eq:dPdnu2}) and
(\ref{eq:Ell}), 
we finally obtain the peak flux,
\begin{equation}
  (\nu F_\nu)_{\rm peak} 
  \approx
  0.0450 f h_{\rm m}^3 \kappa \frac{\mu^2\Omega^4}{c^3}\frac{1}{d^2}
  \label{eq:nuFnu3}
\end{equation}
where $\mu$ denotes the magnetic dipole moment and
\begin{equation}
  f \equiv \frac{n}{0.7} \frac{\varrho_{\rm c}}{0.5\rlc}
           \frac{1}{1.5\kappa}
           \frac{\partial(-\kappa B_z/B)}{\partial(s/\rlc)}
  \label{eq:def_f}
\end{equation}
is close to unity
(see also Zhang \& Cheng~2003).
If we apply $\int_0^\infty x \int_x^\infty K_{5/3}(\xi)d\xi= 1.61$,
we also obtain the integrated flux,
$ \int_0^\infty F_\nu d\nu = 2.36 (\nu F_\nu)_{\rm peak}$.
Note that the factor
$\mu^2\Omega^4/(c^3d^2)$ is proportional to the spin-down flux
(\S~\ref{sec:discussion}). 

It should be emphasized that equation~(\ref{eq:nuFnu3}) estimates 
the upper limit of the phase-{\it averaged} flux.
For example, for an axisymmetric gap in an aligned rotator 
($\inc\!=\!0^\circ$),
the phase-averaged flux equals equation~(\ref{eq:nuFnu3}),
which is constant during the whole NS rotation.
Since equation~(\ref{eq:nuFnu3}) holds particularly well for
an aligned rotator, 
and since the phase-averaged flux decreases with increasing $\inc$
(e.g., MH03 and DHR04),
equation~(\ref{eq:nuFnu3}) gives the upper limit for general $\inc$.
The actual flux will be a few times less than 
equation~(\ref{eq:nuFnu3}), 
because only the field lines in a limited azimuthal range are active,
and because the $\rhoGJ$ at the PC surface decreases
with increasing $\inc$.
In \S~\ref{sec:SG}, we confirm this by comparing with numerical
results.

The difference between the OG and the SG models 
comes into equation~(\ref{eq:nuFnu3}) 
through $h_{\rm m}$, $\kappa$, and the assumed $\mu$.
In \S\S~\ref{sec:OG} and \ref{sec:SG}, 
we apply this equation to these two models,
considering the brightest spin-down-flux pulsar,
the Crab pulsar, assuming $d= 2$~kpc.

\section{Outer-gap Model}
\label{sec:OG}
First, let us apply equation~(\ref{eq:nuFnu3})
to the OG models.
Using the vacuum ($\kappa=1.0$) OG models of 
Cheng et al.~(2000), Takata et al.~(2008) and 
Tang et al.~(2008), which proposed 
$h_{\rm m}\approx 0.11$ 
(i.e., $f \approx 0.11$ in their notation),
we obtain
\begin{eqnarray}
  (\nu F_\nu)_{\rm peak}
  && \!\!\!\!\!\!\!\!\!\!
    \approx 6.58 \times 10^{-4} f \kappa
         \left(\frac{h_{\rm m}}{0.11}\right)^3
         \left(\frac{\mu_{30}}{3.8}\right)^2
  \nonumber\\
  && \hspace*{1.5truecm}
      \mbox{\,MeV\,s}^{-1}\mbox{ cm}^{-2},
  \label{eq:nuFnu4}
\end{eqnarray}
with $\Ell \approx 2.55 \times 10^8 \mbox{ V m}^{-1}$,
$\gamma \approx 2.0 \times 10^7$, 
where $\mu_{30} \equiv \mu/(10^{30}\mbox{\,G\,cm}^3)$.
The $\nu F_\nu$ flux peaks at $1.3h\nu_{\rm c} \approx 4.1$~GeV.
These results are consistent with their vacuum OG models
and with the observed phase-averaged spectrum,
where the photon flux around GeV is dominated by
the primary curvature component rather than the reprocessed
synchrotron-self-Compton one (Takata \& Chang~2007).
Exactly speaking, 
the vacuum OG model is not self-consistent electrodynamically,
because it assumes a vanishing charge density while
it adopts the GJ flux of radiating $e^+$'s.
If one instead adopts the actual $e^+$ flux in a (nearly) vacuum OG,
the predicted $\gamma$-ray flux would be a few orders of magnitude
less than equation~(\ref{eq:nuFnu4}) (left panel of fig.~8 in H06a).
That is, equation~(\ref{eq:nuFnu4}), and hence the traditional OG 
models reproduce observations only phenomenologically.

Secondly, let us consider the non-vacuum OG solution,
which consistently takes the pair production 
in the pulsar magnetosphere into account.
Using $\kappa \approx 0.3$ and $h_{\rm m}\approx 0.14$, we obtain
\begin{eqnarray}
  (\nu F_\nu)_{\rm peak}
  && \!\!\!\!\!\!\!\!\!\!
    \approx 4.06 \times 10^{-4} f \frac{\kappa}{0.3}
         \left(\frac{h_{\rm m}}{0.14}\right)^3
         \left(\frac{\mu_{30}}{3.8}\right)^2
  \nonumber\\
  && \hspace*{1.5truecm}
       \mbox{\,MeV\,s}^{-1}\mbox{\,cm}^{-2},
  \label{eq:nuFnu5}
\end{eqnarray}
which is consistent with the numerical results for Crab
(figs.~8 \& 9 in H08).
In short, the OG model reproduces the observed flux around GeV.

\section{Slot-gap Model}
\label{sec:SG}
Thirdly, let us apply equation~(\ref{eq:nuFnu3}) to the SG model, 
which adopts $h_{\rm m} < 0.05$ 
(i.e., $\Delta \xi_{\rm SG} < 0.05$ in HSDF08's notation)
and $\kappa \approx -(0.1+0.2s/\rlc)$ for the Crab pulsar.
At the outer part ($s\approx 0.6\rlc$),
where emissivity is large, we obtain
\begin{eqnarray}
  (\nu F_\nu)_{\rm peak} 
  && \!\!\!\!\!\!\!\!\!\!
    \approx 6.02 \times 10^{-5} f \frac{\kappa}{0.22}
         \left(\frac{h_{\rm m}}{0.05}\right)^3
         \left(\frac{\mu_{30}}{8}\right)^2
  \nonumber\\
  && \hspace*{1.5truecm}
       \mbox{\,MeV\,s}^{-1}\mbox{\,cm}^{-2},
  \label{eq:nuFnu6}
\end{eqnarray}
with $\vert\Ell\vert \approx 2.4 \times 10^7 \mbox{ V m}^{-1}$ and
$\gamma \approx 1.1 \times 10^7$. 

\begin{figure}
 \epsscale{1.0}
 \plotone{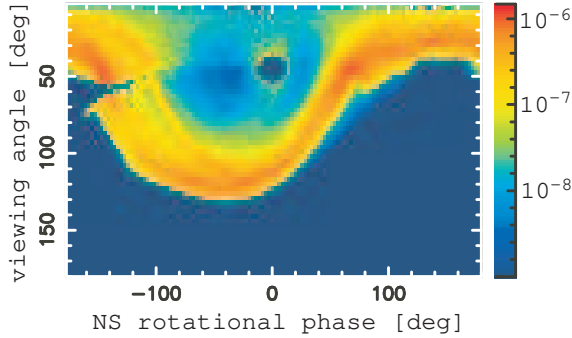} 
\caption{
3-D SG prediction:
Photon flux per phase 
[$\mbox{MeV\,s}^{-1}\mbox{\,cm}^{-2}\mbox{\,deg}^{-1}$]
at distance $d=2$~kpc from the Crab pulsar as a function of 
the viewing angle and the phase.
Only the emission from the SG connected to the north pole 
(i.e., one of the two SGs) is depicted for clarity.
\label{fig:map}
}
\end{figure}

\begin{figure}
 \epsscale{1.0}
 \plotone{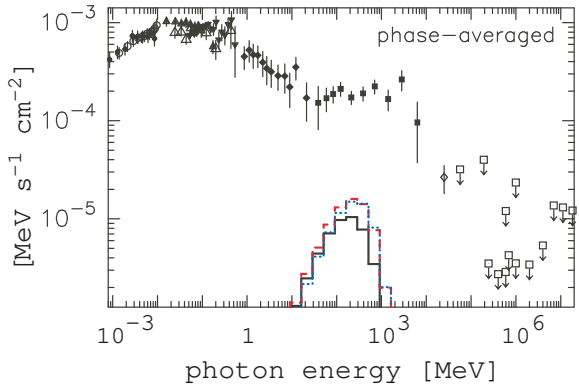} 
\caption{
3-D SG prediction:
Crab phase averaged spectrum at three discrete viewing angles:
$100^\circ$ (solid), $110^\circ$ (dashed), and
$120^\circ$ (dotted).
Emission from both SGs are considered.
For observational data points, see
Kuiper et al.~(2001, and references therein), and
Aliu et al.~(2008) for the flux at 25 GeV.
\label{fig:spectra}
}
\end{figure}

If we adopt the same parameter set 
$h_{\rm m}\!=\!0.04$ and $\mu_{30}\!=\!11$ as HSDF08,
and if we adopt 
$\vert\Ell\vert \approx 7 \times 10^6 \mbox{\,V\,m}^{-1}$,
which is derived from the potential drop 
$0.5 \times 1.3 \times 10^{13}$~V in the higher altitudes SG (HSDF08),
equations~(\ref{eq:nuFnu2}) and (\ref{eq:dPdnu2}) give
$(\nu F_\nu)_{\rm peak} \approx 1.9 \times 10^{-5} 
       \mbox{\,MeV\,s}^{-1}\mbox{\,cm}^{-2}$.
This analytical prediction can be confirmed by a 
numerical computation of the 3-D SG model (H08).
If we adopt the same parameter set as HSDF08
and if we adopt the same $\Ell$ as their equation~(4)
for $\inc\!=\!45^\circ$, we obtain the photon map 
as figure~\ref{fig:map},
which shows a caustic emission from the higher altitudes
by virtue of the exclusion of the strong lower-altitude emission. 
Specifying the observer's viewing angle 
with respect to the rotation axis,
and integrating over the entire NS rotation,
we obtain the phase-averaged spectrum (fig.~\ref{fig:spectra}),
which lies much below the observed value.
Since $\vert\Ell\vert \approx 7 \times 10^6 \mbox{\,V\,m}^{-1}$
gives $\gamma \approx 8.3 \times 10^6$ (eq.~[\ref{eq:termLf}]),
we interpret that HSDF08, who adopted $\gamma \sim 2 \times 10^7$,
overestimated $(\nu F_\nu)_{\rm peak}$ 
by the factor $(2/0.83)^4 \sim 33$.


Equation~(\ref{eq:nuFnu6}) could give the observed flux 
if $\mu_{30}\!>\! 17$.
However, such a large $\mu_{30}$ is not allowed in recent
analyzes of force-free electrodynamics.
For example, Spitkovsky~(2006) derived 
$\dot{E} \!\equiv\! I\Omega\dot{\Omega}
 \!\approx\! (\mu^2 \Omega^4/c^3)(1+\sin^2\inc)$, 
where $I$ denotes the NS moment of inertia
and $\dot{\Omega}$ the temporal derivative of $\Omega$.
Imposing 
$I_{45} \equiv I/(10^{45} \mbox{\,g\,cm}^3) \!<\! 5$
(with $r_\ast \!=\! 16\mbox{\,km}$ and $M \!<\! 2.5 M_\odot$;
 e.g., Lattimer \& Prakash~2000),
we obtain $\mu_{30} \!<\! 3.10 \sqrt{I_{45}} \!<\! 7.07$ for Crab.
Thus, even HSDF08's value, $\mu_{30}=11$, may be a little too large.
In short, 
equation~(\ref{eq:nuFnu6}) gives the conservative upper limit
and the SG model can explain at most 20\% of the observed flux.



\section{Discussion}
\label{sec:discussion}
%

If we assume that the spin down follows the dipole radiation formula,
$\dot{E} \!=\! 2\mu^2\Omega^4\sin^2\inc/(3c^3)$,
equation~(\ref{eq:nuFnu3}) becomes
\begin{equation}
  (\nu F_\nu)_{\rm peak} 
  \approx
  0.0954 f h_{\rm m}^3 \kappa \frac{\dot{E}}{d^2}
           \frac{0.707}{\sin^2\inc},
  \label{eq:nuFnu3c}
\end{equation}
where $\dot{E}/d^2$ denotes the spin-down flux at the Sun.
It was, therefore, natural that the largest spin-down-flux pulsars
were preferentially detected with 
the Energetic Gamma Ray Experiment Telescope (EGRET).
The same tendency can be predicted for Fermi Space Telescope.

For young pulsars, both $h_{\rm m}$ and $\kappa$
are less than unity
(e.g., $h_{\rm m}\approx 0.1$ and $\kappa \approx 0.3$ for Crab)
owing to the copious pair production in the magnetosphere.
As a result, $\int_0^\infty F_\nu d\nu$ becomes much less than
$\dot{E}/d^2$ as demonstrated by H08.
For middle-aged pulsars, on the contrary,
$h_{\rm m}>0.5$ and $\kappa \approx 1$ holds (Hirotani et al.~2003),
leading to an increasing ratio of 
$\int_0^\infty F_\nu d\nu /(\dot{E}/d^2)$ with age.

To explain the observed relationship 
$\int_0^\infty F_\nu d\nu \propto (\dot{E}/d^2)^{0.5}$
of pulsed $\gamma$-ray emissions
(Thompson et al.~1994; Nel et al.~1996),
it is essential to examine the evolution of $h_{\rm m}$ 
with age.
From equation~(\ref{eq:nuFnu3}),
we can at least state that the index $\alpha$ defined by
$\int_0^\infty F_\nu d\nu \propto (\dot{E}/d^2)^\alpha$
becomes less than unity,
because $h_{\rm m}$ increases with decreasing $\dot{E}$,
as discussed just above.
To examine the evolution of $h_{\rm m}$, 
we must solve the screening of $\Ell$ due to the discharge of
the produced pairs in 3-D pulsar magnetospheres.
In subsequent papers, we shall look more carefully into this issue,
by simultaneously solving equation~(\ref{eq:Poisson}),
the Boltzmann equations for $e^\pm$'s, 
and the radiative transfer equation under minimum assumptions.

\acknowledgments
The author is grateful to Drs.
A.~K.~Harding, B.~Rudak (the reviewer), 
K.~S.~Cheng, H.~K.~Chang, and J.~Takata 
for helpful suggestions.
This work is supported by the Theoretical Institute for
Advanced Research in Astrophysics (TIARA) operated under 
Academia Sinica
and the National Science Council Excellence Projects program 
in Taiwan administered through grant number 
NSC 96-2752-M-007-006-PAE.

\acknowledgments

\end{document}